\documentclass[pdftex,final,1p,times]{elsarticle} 
\usepackage{graphicx} 
\usepackage{amssymb} 
\usepackage{amsthm} 
\usepackage{lineno} 

\journal{Nuclear Physics A} 
\begin{document} 

\begin{frontmatter} 
\title{The Long Slow Death of the HBT Puzzle}
\author{Scott Pratt}
\address{Department of Physics and Astronomy,
Michigan State University\\
East Lansing, Michigan 48824, USA}
\date{\today}

\begin{abstract}
At the onset of the RHIC era femtoscopic source sizes inferred from two-particle correlations at RHIC defied description with hydrodynamic models. This failure, which became known as the HBT puzzle, now appears to be solved. The source of the discrepancy appears to be a conspiracy of several factors, each of which contributed to making the evolution of RHIC collisions more explosive. These included pre-equilibrium flow, using a stiffer equation of state and adding viscosity. 
\end{abstract}


\end{frontmatter}

Hydrodynamics represents the foundation of RHIC modeling. The success of ideal hydrodynamics in reproducing elliptic flow and spectra inspired the phrase ``perfect liquid'', as it would appear that the quark-gluon plasma is strongly interacting with perhaps the lowest ratio of viscosity to entropy of any measured substance.  However, these models failed to reproduce femtoscopic source sizes inferred from two-pion correlations (a.k.a. HBT measurements, named after Hanbury-Brown and Twiss who pioneered similar techniques with photons \cite{HanburyBrown:1956pf}). The hydrodynamic models \cite{Soff:2000eh,Teaney:2001av,Hirano:2002hv} also fared much worse than purely microscopic simulations \cite{Petersen:2008gy,Li:2008ge,Humanic:2006sk,Lin:2002gc}, which did not incorporate soft regions in the equation of state. The failure of hydrodynamics, along with the success of what seem to be less plausible models inspired the phrase ``HBT puzzle''. Solving the HBT puzzle would require showing that experimentally inferred source sizes can be reproduced within the context of a hydrodynamic picture, and without destroying agreement with spectra and elliptic flow measurements. Furthermore, the hydrodynamic picture should be interfaced with a cascade to model the breakup stage which is well justified at low density, when interactions between hadrons become binary. If unusual or unjustified physics is added to the model to explain the puzzle, one might simply be exchanging one puzzle for another. For example, one can better reproduce the data with a compact Gaussian profile for the initial energy density \cite{Broniowski:2008vp}, but such a profile is difficult to motivate.

The solution to the puzzle appears to be the conspiracy of several shortcomings of the original hydrodynamic models. Since none of the shortcomings explains more than half of the discrepancy, they were often overlooked as searches tended to focus on finding a single simple explanation of the puzzle. The effects described below are of very different origin, but they all push the dynamics in the same direction, increasing the explosivity. More explosive collisions lead to more compact phase space distributions of outgoing particles. Since HBT analyses effectively measure the size of outgoing phase space clouts, and since these dimensions were over-predicted by the hydrodynamic models, increasing the explosivity improves the agreement with data.

Correlations analyses lead to dimensions of the outgoing phase space distributions. If one looks at particles with asymptotic momenta $p$, their distribution of relative positions is:
\begin{equation}
S_{\bf p}({\bf r})\equiv\frac{\int d^3r_1 d^3r_2~f({\bf p},{\bf r}_1,t)f({\bf p},{\bf r}_2,t)\delta^3({\bf r}_1-{\bf r}_2)}
{\int d^3r_1 d^3r_2~f({\bf p},{\bf r}_1,t)f({\bf p},{\bf r}_2,t)}~.
\end{equation}
This object is referred to as the ``source function'', though it actually refers to the outgoing phase space density $f({\bf p},{\bf r},t)$. It is related to correlation measurements through the Koonin equation,
\begin{eqnarray}\label{eq:koonin}
C({\bf P}={\bf p}_1+{\bf p}_2,{\bf q}=({\bf p_1}-{\bf p}_2)/2)&&\\
\nonumber
&&\hspace*{-80pt}=\int d^3rS_{{\bf P}/2}({\bf r})\left|\phi({\bf q},{\bf r})\right|^2,
\end{eqnarray}
where $\phi$ is the well-understood relative wave function. Rather than fully inverting the measured correlation function to determine the $S_{\bf p}({\bf r})$, most analyses fit to Gaussian forms for the source function,
\begin{equation}
S_{\bf p}({\bf r})\sim \exp\left\{-\frac{x^2}{4R_{\rm out}^2}-\frac{y^2}{4R_{\rm side}^2}
-\frac{z^2}{4R_{\rm long}^2}\right\},
\end{equation} 
and determine the three radius parameters $R_{\rm out}$, $R_{\rm side}$ and $R_{\rm long}$, which are all functions of ${\bf p}$. The longitudinal dimension refers to the beam axis, while the outward axis is in the plane of ${\bf p}$ while being perpendicular to the beam axis. For zero rapidity pairs, or if one boosts along the beam axis so that $p_z=0$, the outward direction is parallel to ${\bf p}$. The sideward dimension is perpendicular to the other two axes. Particles emitted at much different times tend to separate in the outward dimension and lead to large values of $R_{\rm out}$. The exception is when the second particle is emitted further outward a time when the first particle is passing by. Despite this qualifier, the ratio $R_{\rm out}/R_{\rm side}$ is used to gauge the suddenness of the emission. More explosive collisions lead to both smaller $R_{\rm out}/R_{\rm side}$ and smaller values of $R_{\rm long}$. For semi-boost-invariant emission, particles with zero rapidity come only from regions with collective longitudinal velocities that do not exceed the longitudinal thermal velocity $v_{\rm therm}$, i.e., $R_{\rm long}(dv_z/dz)\approx v_{\rm therm}$, where $dv_z/dz$ is the longitudinal velocity gradient. For boost-invariant emission, the velocity gradient equals the inverse emission time. Thus, $R_{\rm long}$ is sensitive to the average emission time, which is smaller for more explosive collisions. It should be emphasized that even though the experimentally determined sources were smaller than hydrodynamic predictions, they are still far larger than the initial spatial extent of the reaction zone. Longitudinal sizes approach 10 fm, whereas the initial nuclei are compressed to a tenth of a fm, and the sideward dimension are also double the initial transverse size. Given that the phase space distribution dimensions refer to only those particles with a specific momentum, the overall source sizes are significantly larger than the sizes of the sub-regions quoted in HBT analyses. This is perhaps the most direct evidence of the dramatic expansion of the interacting fireball in RHIC collisions and underscores the folly of ignoring longitudinal transverse expansion.

\begin{figure}
\centerline{\includegraphics[width=0.7\textwidth]{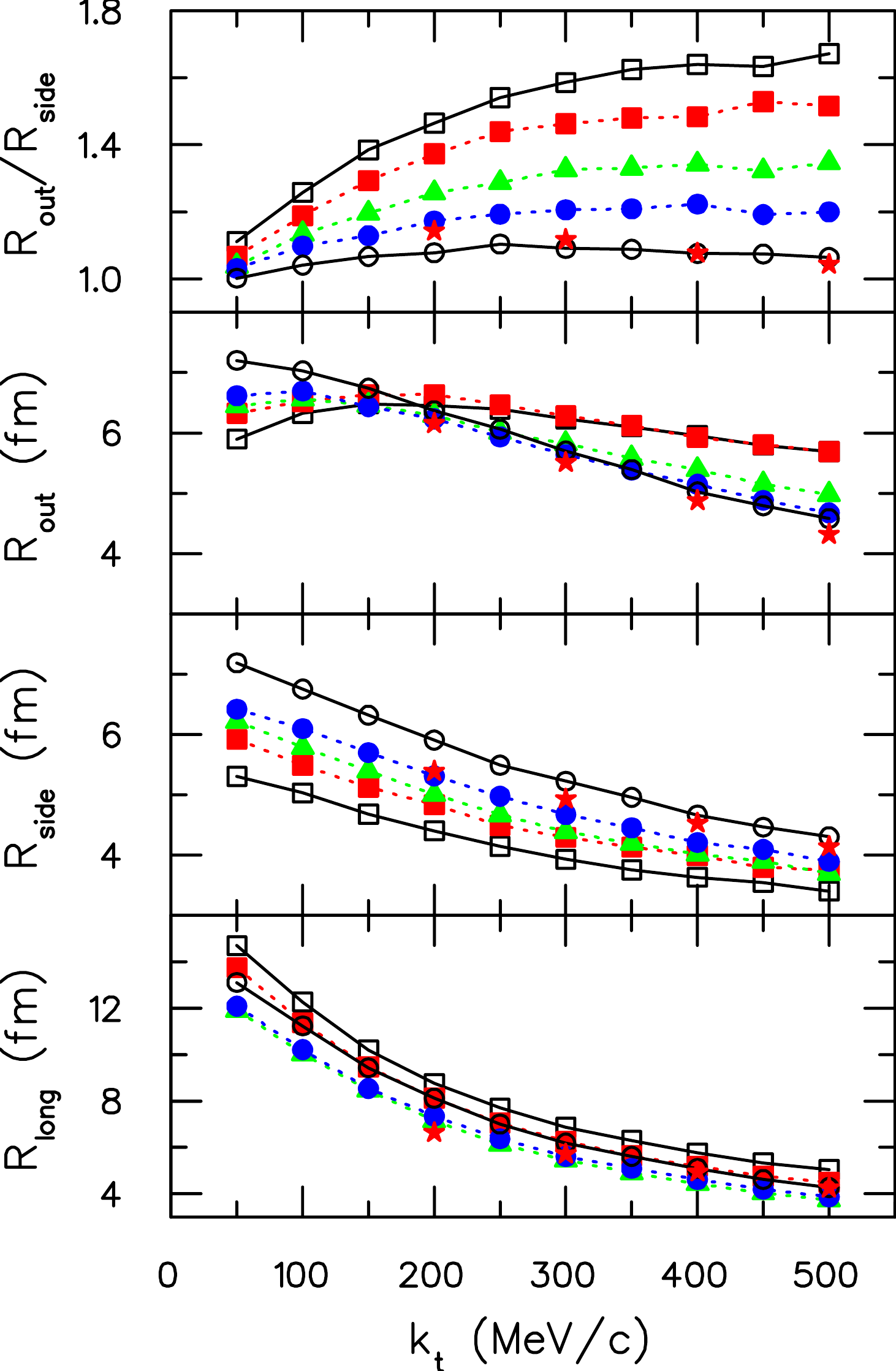}}
\caption{\label{fig:Rosl_everything}(color online)
Gaussian radii for three dimensions: $R_{\rm out}$, $R_{\rm side}$ and $R_{\rm long}$. Data from STAR (red stars) are poorly fit by a model with a first-order phase transition, no pre-thermal flow, and no viscosity (open black squares and solid line). Data are better reproduced after including all the features (open circles, solid black line). The incremental improvements (open colored symbols and dashed lines) are of similar strength: initial flow (red squares), stiffer equation of state (green triangles), viscosity (blue circles). Incorporating a more compact initial profile represents the final feature.}
\end{figure}

After nearly a decade of searching for {\it THE} solution to the HBT puzzle, it now appears that the explanation comes from the conspiracy of multiple features which were missing or misrepresented in the early models. None of these causes represents more than half of the original discrepancy by themselves, but after including all the features (or at least the first three below) the models reproduce source sizes within the systematic error of the models (often stated as being close to 10\%). Each of these effects pushes the resulting source sizes in the same direction. To illustrate each improvement, Figure \ref{fig:Rosl_everything} presents extracted source sizes for a model with none of the features along with results for each feature added incrementally. The model is a viscous hydrodynamic calculation coupled to a hadronic cascade, which models the evolution once temperatures fall below 170 MeV. By assuming both boost-invariant and azimuthal symmetries the model needs only to consider a one-dimensional radial mesh, which makes for rapid calculations. After each feature is added the original energy density is rescaled to match the final experimental $dN_{\rm ch}/d\eta$. The speed of the calculations makes it convenient to analyze the impact of the features independently, but unfortunately, elliptic flow analyses are precluded by the reduced dimensionality. As a benchmark (open squares in Fig. \ref{fig:Rosl_everything}) a calculation was performed with none of the additional features listed below. It fails in very similar ways, both qualitatively and quantitatively, to previous hydrodynamic calculations missing these features. The improvements below are then added incrementally.
\begin{enumerate}
\item Adding pre-equilibrium flow. First generation hydrodynamic models initialized the evolution at a thermalization time, usually chosen between 0.5 and 1.0 fm/$c$. However, flow can and should develop even before the matter is thermalized. The importance of pre-thermalized acceleration for HBT studies has been demonstrated in several studies during the last few years \cite{Gyulassy:2007zz,Li:2008ge,Broniowski:2008vp}, and has been investigated in greater detail in regards to other observables \cite{Kolb:2002ve,Heinz:2002rs,Krasnitz:2002ng}. Remarkably, this flow is of similar strength for a wide variety of pictures of the initial state \cite{Vredevoogd:2008id}. The initial impulse of flow results in particle emission that is both earlier and more sudden. The closed red squares in Fig. \ref{fig:Rosl_everything} show how this improves the calculation relative to the benchmark calculation, by comparing calculations where the hydrodynamic expansion was started at 0.1 fm/$c$ vs. 1.0 fm/$c$.
\item The equations of state used in many of the early studies was often too soft. In most cases a first order phase transition was assumed. For a first order transition the pressure stays fixed for a range of energy densities, whose range defines the latent heat of the transition. This is inconsistent with lattice calculations which only show a dip in the speed of sound, $c_s^2=dP/d\epsilon$. Higher pressures lead to more explosive collisions, and the improvement (closed green triangles vs. closed red squares) in Fig. \ref{fig:Rosl_everything} is substantial. The calculations used simplified forms for the equation of state. For temperatures below 170 MeV, or equivalently for energy densities below $\epsilon_h$, the equation of state was that of a hadron gas. After that point $dP/d\epsilon=c_s^2$ was fixed for a range of energy densities, $L$, chosen to be 1.6 GeV/fm$^3$ in the default calculation and 800 MeV/fm$^3$ for the stiffer case. The speed of sound was chosen to be zero in the default calculation, i.e., a first-order transition, and chosen to be 0.1 for the more realistic case. Once the energy density exceeded $\epsilon_h+L$, the speed of sound was chosen to be $c_s^2=0.3$, consistent with lattice calculations.
\item Shear viscosity leads to anisotropic stress energy tensors. Since the expansion is stronger along the the beam axis, the transverse pressure is increased relative to the longitudinal pressure. This also increases the explosivity \cite{Teaney:2003kp,Romatschke:2007mq,Pratt:2008sz}. For energy densities above $\epsilon_h+L$ the shear was raised from $4\pi\eta/s=0$ to $2$. For lower energy densities the shear viscosity was set equal to that of a hadron gas, and for the intermediate range, $\epsilon_h<\epsilon<\epsilon_h+L$, the viscosity varied linearly with energy density, effectively as a linear interpolation between $\epsilon_h$ and $\epsilon_h+L$. A Bulk viscosity was also added, see \cite{Pratt:2008sz}, but had little effect on the source sizes though it reduced radial flow. The change in the source sizes (filled green triangles to filled blue circles in Fig. \ref{fig:Rosl_everything}) is again substantial and along with the two previous improvements brings the result with the systematic error of the data. 
\item Using a more compact or more Gaussian initial energy density profiles also increases the explosivity. Gaussian profiles \cite{Broniowski:2008vp} lead to nearly perfectly linear velocity profiles, as opposed to Wood-Saxon profiles which have higher density gradients near the surface. Having the profiles steeper at the surface allows the early emission to gain an advantage over the later emission from the core, which leads to larger values of $R_{\rm out}/R_{\rm long}$. Although some color glass arguments can lead to more compact profiles, there are no good motivations for the large effects of \cite{Broniowski:2008vp} which require Gaussian shapes. However, even modestly more compact profiles lead to non-negligible changes in the source radii. The effects of employing a color-glass inspired \cite{Drescher:2007cd} profile, vs. a more standard wounded nucleon picture (open circles vs. filled blue circles) is apparent in Fig. \ref{fig:Rosl_everything}.
\item A final improvement to the model comes from enacting a more accurate treatment of the two-pion wave function for the relative wave applied to the Koonin equation, Eq. (\ref{eq:koonin}). Most theoretical treatments have ignored the strong and Coulomb interactions between the pions as they are compared to experimental results for which the effect of Coulomb interactions is largely divided away with what is known as the Bowler-Sinukov procedure \cite{Bowler:1991vx,Sinyukov:1998fc}. A truer comparison involves calculating the theoretical correlation functions with the Coulomb and strong interaction, then performing the same Bowler-Sinyukov procedure on the theoretical correlations. One then compares the resulting correlation functions to what one would obtain from Gaussian sources assuming no such complicated interactions, consistent with how the data are treated. The more accurate procedure results in source sizes that change only at the level of a percent or two \cite{Pratt:2008sz}, but again, the corrections are in the same direction as those listed above. 
\end{enumerate}
\begin{figure}

\centerline{\includegraphics[width=0.6\textwidth]{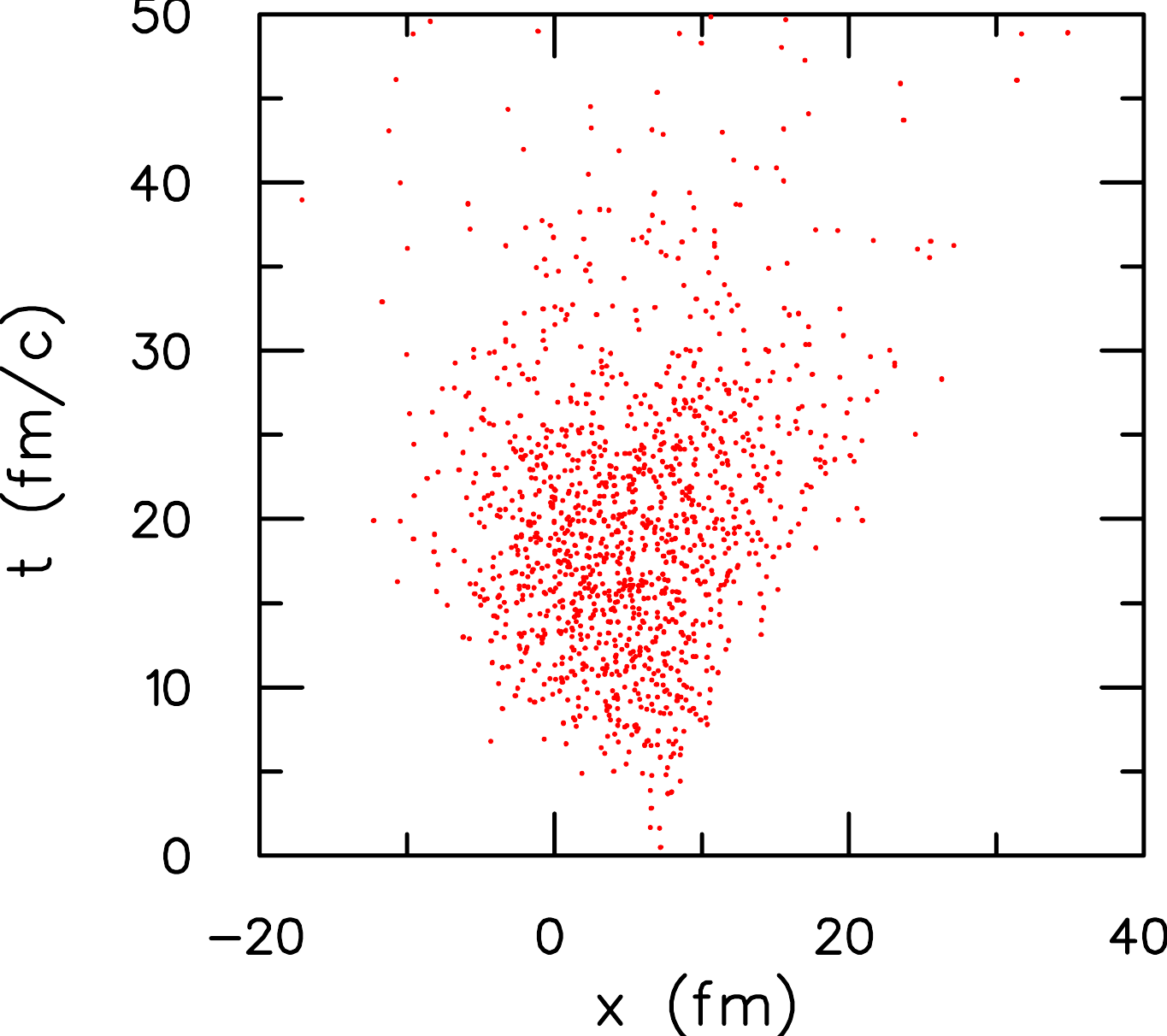}}
\caption{\label{fig:xyzt}(color online)
Final emission positions and times for particles with transverse momentum of 300 MeV/$c$ along the $x$ (outward) axis. Emission is surface dominated and has a modestly positive correlation between $x$ and $t$.}
\end{figure}
A second aspect of the HBT puzzle concerns fits with blast-wave models, which are based on a picture of thermal emission from a collectively expanding source, parameterized by a breakup temperature $T$, an outer radius $R$, a breakup time $\tau$, a linearly rising transverse collective velocity with a maximum, $v_{\rm max}$, and an emission duration $\Delta\tau$. These fits revealed outer radii near a dozen fm, with emission confined to within a few fm/$c$ of $\tau=10$ fm/$c$ \cite{Retiere:2003kf,Kisiel:2006is}. Even though these radii are noticably larger than the HBT radii, which reflect only a subset of the overall size, the parameters suggest an unphysically high breakup density. Combining such densities with standard hadronic cross sections suggests mean free paths of 1-2 fm. However, breakup should not occur until the mean free path is of order of the system size. Figure \ref{fig:xyzt} shows the outward coordinate $x$ and the time $\tau$ at which emission occured for the final model in Fig. \ref{fig:Rosl_everything}, and are similar to what was seen in \cite{Lin:2002gc}. Points are shown only for particles emitted with momentum $p_x=300$ MeV/$c$. Emission comes mostly from within a few fm of the surface, and there exists a modestly positive correlation between $x$ and $t$, as the average $x$ for the emission points moves outward at approximately a tenth the speed of light. A positive correlation prevents those particles produced at later times from being strongly separated from those emitted earlier, which leads to smaller outward sizes of the outgoing phase space packet. Other physics elements of the breakup are also missing from blast wave pictures, such as the differential cooling and collective flow of protons and pions once they lose equilibrium \cite{Pratt:1998gt}. This emphasizes the importance of using realistic dynamical models to compare to femtoscopic data and underscores the limits of parametric fits.

\begin{figure}
\centerline{\includegraphics[width=0.6\textwidth]{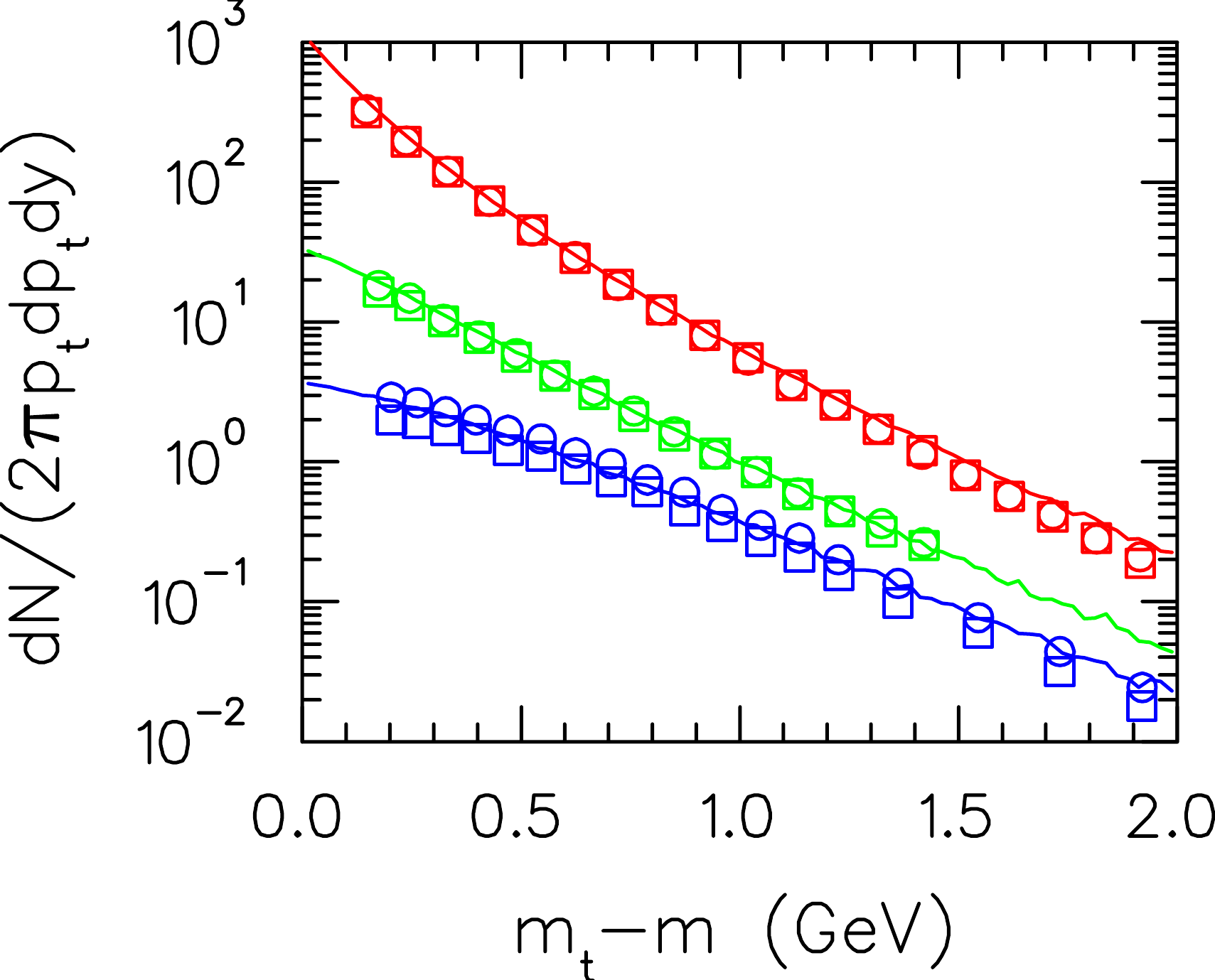}}
\caption{\label{fig:spectra}(color online)
Model spectra for pions, kaons and $p/\bar{p}$ and spectra (top to bottom, model represented by lines) are compared to PHENIX data \cite{Adler:2003cb}. Positive/negative species are represented by circles/squares.}
\end{figure}
The features included above also affect spectra and elliptic flow observables. Unfortunately, the symmetries applied in the model used for this study precludes an elliptic flow analysis. As shown in Fig. \ref{fig:spectra}, the model does match experimental spectra despite the added explosivity (although the baryon yields were significantly off). Some of the explosivity might have been balanced by the bulk viscosity, which had little effect on the HBT radii. A more detailed study should be undertaken to see how the various features listed above impact spectra. In addition to spectra, elliptic anisotropies and HBT, the entire host of low-$p_t$ observables should be simultaneously addressed to validate a model. This includes more detailed aspects of femtoscopic measurements: radii with respect to the reaction plane \cite{Adams:2003ra}, correlations of other species, especially non-identical particles \cite{Kisiel:2004it,Adams:2003qa}, and non-Gaussian features of the source function \cite{Panitkin:2001qb}. Even if all these data are reproduced, it does not fully validate the model. That would require an ambitious statistical analysis of the set of model parameters and assumptions, similar to what has been applied to cosmic microwave background analyses \cite{Habib:2007ca}. Although these goals require significant effort in the coming years, the current analysis has eliminated any puzzle about femtoscopy for the time being, as the experimental radii appear to be satisfactorily described within a rather standard theoretical picture of RHIC collision dynamics.

\section*{Acknowledgments} The author is indebted to all the theorists who provided the ideas that led to the solution of this puzzle, as well to the experimentalists who have provided the beautifully analyzed data. This work was generously supported by the U.S. Department of Energy Office of Science through grant DE-FG02-03ER41259.

\end{document}